# Bit Error Rate is Convex at High SNR

Sergey Loyka

SITE
University of Ottawa
Ottawa, K1N 6N5, Canada
e-mail: sergey.loyka@ieee.org

Victoria Kostina

Department of Electrical Engineering
Princeton University
Princeton, NJ, 08544, USA
e-mail: vkostina@princeton.edu.

Francois Gagnon

Department of Electrical Engineering
Ecole de Technologie Superieure
Montreal, H3C 1K3, Canada
e-mail: francois.gagnon@etsmtl.ca

**Abstract— Motivated by a wide-spread use of convex optimization techniques, convexity properties of bit error rate of the maximum likelihood detector operating in the AWGN channel are studied for arbitrary constellations and bit mappings, which may also include coding under maximum-likelihood decoding. Under this generic setting, the pairwise probability of error and bit error rate are shown to be convex functions of the SNR in the high SNR regime with explicitly-determined boundary. The bit error rate is also shown to be a convex function of the noise power in the low noise/high SNR regime.**

## I. INTRODUCTION

Optimization problems of various kinds simplify significantly if the goal and constraint functions involved are convex. Indeed, a convex optimization problem has a unique global solution, which can be found either analytically or, with a reasonable effort, by several efficient numerical methods; its numerical complexity grows only moderately with the problem dimensionality and required accuracy; convergence rates and required step size can be estimated in advance; there are powerful analytical tools that can be used to attack a problem and that provide insights into such problems even if solutions, either analytical or numerical, are not found yet [1][2]. Contrary to this, not only generic nonlinear optimization problems do not possess these features, they are not solvable numerically, i.e. their complexity grows prohibitively fast with problem dimensionality and required accuracy [2]. Thus, there is a great advantage in formulating or at least in approximating an optimization problem as a convex one.

In the world of digital communications, one of the major performance measures is either symbol or bit error rate (SER or BER). Consequently, when an optimization of a communication system is performed, either SER or BER often appears as goal or constraint functions. Examples include optimum power/rate allocation in spatial multiplexing systems (BLAST) [3], optimum power/time sharing for a transmitter and a jammer [4], rate allocation or precoding in multicarrier (OFDM) systems [5], optimum equalization [6], optimum multiuser detection [7], and joint Tx-Rx beamforming (precoding-decoding) in MIMO systems [8]. Symbol and bit error rates of the maximum likelihood (ML) detector have been extensively studied and a large number of exact or approximate analytical results are available for various modulation formats, for both non-fading and fading AWGN channels [9][10]. On the other hand, convexity properties of error rates are not understood so well, especially for constellations of complicated geometry, large dimensionality or when coding is used. Results in this area are scarce. Many known closed-form error rate expressions can be verified by differentiation to be convex, but this approach does not provide any generic conclusions. Convexity properties for binary modulations have been studied in-depth in [4], including applications to transmitter and jammer optimizations, and were later extended to arbitrary multidimensional constellations in [11][12] in terms of the SER under maximum-likelihood detection. A log-concavity property of the SER as a function of the SNR [dB] for the uniform square-grid constellations has been established by Conti et al [13].

Unfortunately, convexity of SER does not say anything in general about convexity of the BER, since the latter depends on pairwise probabilities of error (PEP) and not on the SER [14]. Since the BER is an important performance indicator and thus appears as an objective in many optimization problems, we study its convexity in the present paper using a generic geometrical framework developed in [11][12]. Our setting is generic enough so that the results apply to constellations of arbitrary order, shape and dimensionality, which may also include coding

First, we establish convexity properties of the PEP as a function of SNR: it is convex at high SNR regime, concave at the low one, and has an odd number of inflection points in-between. Based on this, convexity of the BER at high SNR is established for arbitrary constellation and coding. Thus, this property is a consequence of Gaussian noise density and maximum likelihood detection rather than particular constellation or coding technique. We also show that the BER is a convex function of the noise power in the small noise/high SNR mode.

## II. SYSTEM MODEL

The standard baseband discrete-time system model with an AWGN channel, which includes matched filtering and sampling, is

$$\mathbf{r} = \mathbf{s} + \boldsymbol{\xi} \qquad (1)$$

where $\mathbf{s}$ and $\mathbf{r}$ are $n$-dimensional vectors representing the Tx and Rx symbols respectively, $\mathbf{s} \in \{\mathbf{s}_1, \mathbf{s}_2, ..., \mathbf{s}_M\}$, a set of $M$ constellation points, $\boldsymbol{\xi}$ is the additive white Gaussian noise (AWGN), $\boldsymbol{\xi} \sim \mathcal{N}(\mathbf{0}, \sigma_0^2 \mathbf{I})$, whose probability density function (PDF) is

$$p_\xi(\mathbf{x}) = \left(2\pi\sigma_0^2\right)^{-n/2} e^{-|\mathbf{x}|^2/2\sigma_0^2} \qquad (2)$$

where $\sigma_0^2$ is the noise variance per dimension, and $n$ is the constellation dimensionality; lower case bold letters denote vectors, bold capitals denote matrices, $x_i$ denotes i-th component of $\mathbf{x}$, $|\mathbf{x}|$ denotes L$_2$ norm of $\mathbf{x}$, $|\mathbf{x}| = \sqrt{\mathbf{x}^T \mathbf{x}}$, where the superscript $T$ denotes transpose, $\mathbf{x}_i$ denotes i-th vector. The average (over the constellation points) SNR is defined as $\gamma = 1/\sigma_0^2$, which implies the appropriate normalization, $\frac{1}{M}\sum_{i=1}^{M}|\mathbf{s}_i|^2 = 1$.

Consider the maximum likelihood detector, which is equivalent to the minimum distance one in the AWGN channel, $\hat{\mathbf{s}} = \arg\min_{\mathbf{s}_i}|\mathbf{r} - \mathbf{s}_i|$. The probability of symbol error $P_{ei}$ given that $\mathbf{s} = \mathbf{s}_i$ was transmitted is $P_{ei} = \Pr[\hat{\mathbf{s}} \neq \mathbf{s}_i | \mathbf{s} = \mathbf{s}_i] = 1 - P_{ci}$, where $P_{ci}$ is the probability of correct decision. The SER averaged over all constellation points is $P_e = \sum_{i=1}^{M} P_{ei} \Pr[\mathbf{s} = \mathbf{s}_i] = 1 - P_c$. $P_{ei}$ can be expressed as

$$P_{ei} = 1 - \int_{\Omega_i} p_\xi(\mathbf{x}) d\mathbf{x} \quad (3)$$

where $\Omega_i$ is the decision region (Voronoi region), and $\mathbf{s}_i$ corresponds to $\mathbf{x} = 0$, i.e. the origin is shifted for convenience to the constellation point $\mathbf{s}_i$. $\Omega_i$ can be expressed as a convex polyhedron [1],

$$\Omega_i = \{\mathbf{x} | \mathbf{A}\mathbf{x} \leq \mathbf{b}\}, \quad \mathbf{a}_j^T = \frac{(\mathbf{s}_j - \mathbf{s}_i)}{|\mathbf{s}_j - \mathbf{s}_i|}, \quad b_j = \frac{1}{2}|\mathbf{s}_j - \mathbf{s}_i| \quad (4)$$

where $\mathbf{a}_j^T$ denotes j-th row of $\mathbf{A}$, and the inequality in (4) is applied component-wise. Clearly, $P_{ei}$ and $P_{ci}$ posses the opposite convexity properties.

Another important performance indicator is the pairwise probability of error (PEP) i.e. a probability $\Pr\{\mathbf{s}_i \to \mathbf{s}_j\} = \Pr[\hat{\mathbf{s}} = \mathbf{s}_j | \mathbf{s} = \mathbf{s}_i]$ to decide in favor of $\mathbf{s}_j$ given that $\mathbf{s}_i$, $i \neq j$, was transmitted, which can be expressed as

$$\Pr\{\mathbf{s}_i \to \mathbf{s}_j\} = \int_{\Omega_j} p_\xi(\mathbf{x}) d\mathbf{x} \quad (5)$$

where $\Omega_j$ is the decision region for $\mathbf{s}_j$ when the reference frame is centered at $\mathbf{s}_i$. The SER can now be expressed as

$$P_{ei} = \sum_{j \neq i} \Pr\{\mathbf{s}_i \to \mathbf{s}_j\} \quad (6)$$

and the BER can be expressed as a positive linear combination of PEPs [14]

$$\text{BER} = \sum_{i=1}^{M} \sum_{j \neq i} \frac{h_{ij}}{\log_2 M} \Pr\{\mathbf{s} = \mathbf{s}_i\} \Pr\{\mathbf{s}_i \to \mathbf{s}_j\} \quad (7)$$

where $h_{ij}$ is the Hamming distance between binary sequences representing $\mathbf{s}_i$ and $\mathbf{s}_j$.

Note that the model and error rate expressions we are using are generic enough to apply to arbitrary constellations, which may also include coding under maximum-likelihood decoding (codewords are considered as points of an extended constellation). We now proceed to establish convexity properties of error rates in this generic setting.

## III. CONVEXITY OF SYMBOL ERROR RATES

Convexity properties of symbol error rates for arbitrary constellations in the SNR and noise power have been established in [11][12] and are summarized below for completeness and comparison purpose.

**Theorem 1 (Theorem 1 and 2 in [11])**: The SER $P_e$ is a convex function of the SNR $\gamma$ for any constellation (which may also include coding) if $n \leq 2$,

$$d^2 P_e / d\gamma^2 = P_{e|\gamma}'' > 0 \quad (8)$$

For $n > 2$, the following convexity properties hold:
- $P_{ei}$ is convex in the large SNR mode,

$$\gamma \geq (n + \sqrt{2n})/d_{\min,i}^2 \quad (9)$$

where $d_{\min,i}$ is the minimum distance from $\mathbf{s}_i$ to its decision region boundary,

- $P_{ei}$ is concave in the small SNR mode,

$$\gamma \leq (n - \sqrt{2n})/d_{\max,i}^2 \quad (10)$$

where $d_{\max,i}$ is the maximum distance from $\mathbf{s}_i$ to its decision region boundary,

- there are an odd number of inflection points, $P_{ci|\gamma}'' = P_{ei|\gamma}'' = 0$, in the intermediate SNR mode,

$$(n - \sqrt{2n})/d_{\max,i}^2 \leq \gamma \leq (n + \sqrt{2n})/d_{\min,i}^2 \quad (11)$$

- the SER $P_e$ is convex at high SNR,

$$\gamma \geq (n + \sqrt{2n})/d_{\min}^2 \quad (12)$$

where $d_{\min} = \min_i \{d_{\min,i}\}$ is the minimum distance to decision region boundary in the constellation.

**Theorem 2 (Theorem 4 in [11]):** Symbol error rates have the following convexity properties in the noise power $P_N = \sigma_0^2$, for any $n$ and constellation geometry,
- $P_{ei}$ is concave in the large noise mode,

$$P_N \geq d_{\max,i}^2 \left(n + 2 - \sqrt{2(n+2)}\right)^{-1} \quad (13)$$

- $P_{ei}$ is convex in the small noise mode,

$$P_N \leq d_{\min,i}^2 \left(n + 2 + \sqrt{2(n+2)}\right)^{-1} \quad (14)$$

- there are an odd number of inflection points for intermediate noise power,

$$d_{\min,i}^2 \left(n + 2 + \sqrt{2(n+2)}\right)^{-1} \leq P_N \leq d_{\max,i}^2 \left(n + 2 - \sqrt{2(n+2)}\right)^{-1} \quad (15)$$

- the SER $P_e$ is convex in the small noise/high SNR mode,

$$P_N \leq d_{\min}^2 \left(n + 2 + \sqrt{2(n+2)}\right)^{-1} \quad (16)$$

While the convexity properties above are important for many optimization problems, they do not lend any conclusions about convexity of the BER, since the latter is not directly related to $P_e$ or $P_{ei}$ in general. While, in some cases, the BER can be expressed as linear combination of $P_{ei}$, there are positive and negative terms so that no conclusion about convexity can be made in this case either. On the other hand, the BER can be expressed as a positive linear combination of pairwise probabilities of error so that the convexity of the latter implies the convexity of the former. Thus, we study below the convexity property of the PEP, from which the convexity property of the BER will follow.

## IV. CONVEXITY OF PAIRWISE PROBABILITY OF ERROR

In many cases, it is a pairwise error probability that is a key point in the analysis (e.g. for constructing a union bound and other performance metrics). Furthermore, it is also a basic building block for the BER in (7), so that we establish its convexity property first.

**Theorem 3**:

a) The pairwise error probability $\Pr\{\mathbf{s}_i \to \mathbf{s}_j\}$ is a convex function of the SNR at the high SNR region, $\gamma \geq (n+\sqrt{2n})/d_{\min,i}^2$, for any $n$;

b) for $n=1,2$, it is concave at the low SNR region, $\gamma \leq (n+\sqrt{2n})/(d_{ij}+d_{\max,j})^2$, where $d_{ij}=|\mathbf{s}_i-\mathbf{s}_j|$ is the distance between $\mathbf{s}_i$ and $\mathbf{s}_j$, and there is an odd number of inflection points, $\Pr\{\mathbf{s}_i \to \mathbf{s}_j\}'' = 0$, in the intermediate SNR mode,

$$(n+\sqrt{2n})/(d_{ij}+d_{\max,j})^2 \leq \gamma \leq (n+\sqrt{2n})/d_{\min,i}^2 \quad (17)$$

c) for $n>2$, the PEP is convex at the low SNR region, $\gamma \leq (n-\sqrt{2n})/(d_{ij}+d_{\max,j})^2$, and there is an even number of inflection points in-between,

$$(n-\sqrt{2n})/(d_{ij}+d_{\max,j})^2 \leq \gamma \leq (n+\sqrt{2n})/d_{\min,i}^2$$

**Proof:** See Appendix.

We note that Theorem 3(a) is stronger than Theorem 1 at the high SNR region since the latter follows from the former but the opposite is not always true (as the other SNR ranges in Theorem 3 above indicate). Unlike the SER, the pairwise error probability can be concave at low SNR even for $n=1,2$.

Since Theorem 3 holds for any constellation and bit mapping, it follows that the convexity property of the PEP at high SNR is a consequence of Gaussian noise density rather than particular modulation/coding used, where the latter determines only the SNR threshold.

## V. CONVEXITY OF THE BER AT HIGH SNR

We are now in a position to establish the main result of this paper.

**Theorem 4**: The BER is a convex function of the SNR, for any constellation and bit mapping, which may also include coding under maximum-likelihood decoding, at the high SNR regime,

$$\gamma \geq (n+\sqrt{2n})/d_{\min}^2, \quad (18)$$

where $d_{\min} = \min_i \{d_{\min,i}\}$ is the minimum distance to the boundary in the constellation.

**Proof:** Using the relationship between the BER and the pairwise error probabilities in (7) and observing that a positive linear combination of convex functions is convex. Q.E.D.

We remark that the condition in (18) guarantees the convexity of all PEP, BER and SER. In some cases (Gray encoding and when nearest neighbor errors dominate), the BER is approximated as $\text{SER}/\log_2 M$, so that it inherits the same convexity properties as in Theorems 1 and 2 above.

## VI. CONVEXITY OF THE PEP AND BER IN NOISE POWER

In a jammer optimization problem, it is convexity properties in noise power that are important [4]. Motivated by this fact, we study below convexity of the PEP and BER in the noise power.

**Theorem 5:** The PEP $\Pr\{\mathbf{s}_i \to \mathbf{s}_j\}$ is a convex function of the noise power $P_N = \sigma_0^2$, for any $n$, in the small noise/high SNR mode,

$$P_N \leq d_{\min,i}^2 \left(n+2+\sqrt{2(n+2)}\right)^{-1} \quad (19)$$

and in the large noise/low SNR mode,

$$P_N \geq (d_{ij}+d_{\max,j})^2 \left(n+2-\sqrt{2(n+2)}\right)^{-1} \quad (20)$$

**Proof:** See Appendix.

Based on this Theorem, the following convexity property of the BER is established.

**Corollary 5.1**: For any constellation geometry and dimensionality, which may also include coding under ML decoding, the BER is a convex function of the noise power in the small noise/high SNR mode:

$$P_N \leq d_{\min}^2 \left(n+2+\sqrt{2(n+2)}\right)^{-1} \quad (21)$$

where specifics of the constellation/code determine only the upper bound in (21).

## VIII. APPENDIX

**Proof of Theorem 3:** The pairwise probability of error $P_{ij} = \Pr\{\mathbf{s}_i \to \mathbf{s}_j\}$ can be presented as

$$P_{ij} = \int_{\Omega_j} p_\xi(\mathbf{x}) d\mathbf{x} \qquad (22)$$

where $\Omega_j$ is the decision region for $\mathbf{s}_j$ when the reference frame is centered at $\mathbf{s}_i$. Its second derivative in the SNR is

$$P_{ij}'' = \int_{\Omega_j} \frac{d^2 p_\xi(\mathbf{x})}{d\gamma^2} d\mathbf{x} \qquad (23)$$

where the derivative is

$$\frac{d^2 p_\xi(\mathbf{x})}{d\gamma^2} = \frac{1}{4} \left( \frac{\gamma}{2\pi} \right)^{n/2} e^{-\gamma|\mathbf{x}|^2/2} f\left(|\mathbf{x}|^2\right) \qquad (24)$$

and $f(t) = (t - \alpha_1/\gamma)(t - \alpha_2/\gamma)$, $\alpha_1 = n + \sqrt{2n} > 0$, $\alpha_2 = n - \sqrt{2n} < \alpha_1$. Consider three different cases.

(i) If $d_{\min,i}^2 \geq \alpha_1/\gamma$, where $d_{\min,i} = \min_j(b_j)$ is the minimum distance from the origin to the boundary of $\Omega_i$, then $f(|\mathbf{x}|^2) \geq 0 \ \forall \mathbf{x} \in \Omega_j$ so that the integral in (23) is clearly positive since the integrand is non-negative everywhere in the integration region and positive in some parts of it. Fig. 1 illustrates this case. This is a high SNR mode since $\gamma \geq \alpha_1/d_{\min,i}^2$.

(ii) If $(d_{ij} + d_{\max,j})^2 \leq \alpha_1/\gamma$ and $n = 1, 2$, where $d_{\max,j}$ is the maximum distance from the center of $\Omega_j$ to its boundary, then $f(|\mathbf{x}|^2) \leq 0 \ \forall \mathbf{x} \in \Omega_j$ so that the integral in (23) is clearly negative and the result follows. Fig. 2 illustrates this case. This is a low-SNR mode since $\gamma \leq \alpha_1/(d_{ij} + d_{\max,j})^2$. An odd number of inflection points in Theorem 3(b) follows from the continuity argument ($P_{ij}''$ is a continuous function of the SNR).

(iii) Part (c) follows from the same argument as in (ii). Q.E.D.

**Proof of Theorem 5:** follows the same geometric technique as for Theorem 3. 2nd derivative of the PEP in the noise power can be expressed as

$$\frac{d^2 P_{ij}}{dP_N^2} = \int_{\Omega_j} \frac{d^2 p_\xi(\mathbf{x})}{P_N^2} d\mathbf{x} \qquad (25)$$

where

$$\frac{d^2 p_\xi(\mathbf{x})}{dP_N^2} = \frac{1}{4P_N^4} \left( \frac{1}{2\pi P_N} \right)^{\frac{n}{2}} e^{-\frac{|\mathbf{x}|^2}{2P_N}} f^*\left(|\mathbf{x}|^2\right)$$

$$f^*(t) = (t - \beta_1 P_N)(t - \beta_2 P_N), \qquad (26)$$

$$\beta_1 = n + 2 + \sqrt{2(n+2)}, \quad \beta_2 = n + 2 - \sqrt{2(n+2)}$$

and $\beta_1 > \beta_2 > 0$. Since $f^*(t)$ has the same structure as $f(t)$ in (24), the proof follows the same steps. In particular, if $d_{\min,i}^2 \geq \beta_1 P_N$, then $d^2 p_\xi / dP_N^2 > 0 \ \forall \mathbf{x} \in \Omega_j$ so that the integral in (25) is clearly positive. The other case is proved in a similar way. Q.E.D.

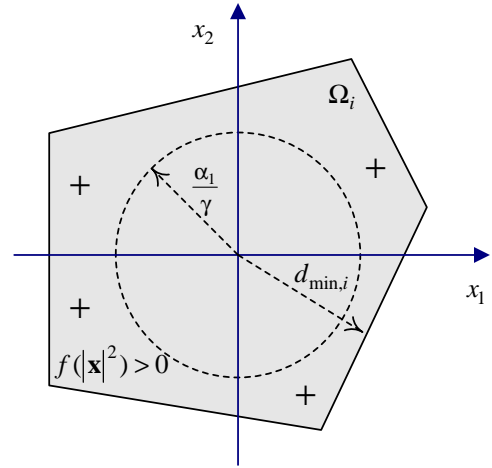

**Fig. 1.** Two-dimensional illustration of the problem geometry for Case 1. The decision region $\Omega_i$ is shaded. $f(|\mathbf{x}|^2)$ has a sign as indicated by "+" and "-".

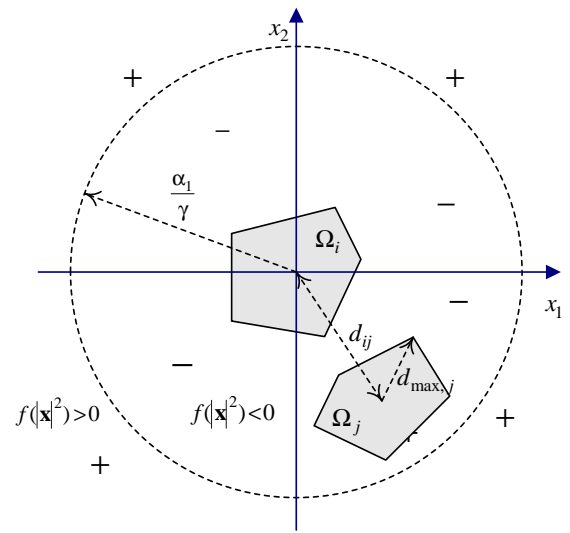

**Fig. 2.** Two-dimentional illustration of the problem geometry for Case 2.